# Progress and Prospects In Magnetic Topological Materials

B. Andrei Bernevig[1], Claudia Felser[2], and Haim Beidenkopf[3]

[1] *Department of Physics, Princeton University, Princeton, New Jersey 08544, USA*

[2] *Max Plank Institute Chemical Physics of Solids, Dresden, Germany*

[3] *Department of Condensed Matter Physics, Weizmann Institute of Science, Rehovot 7610001, Israel*

Magnetic topological materials represent a class of compounds whose properties are strongly influenced by the topology of the electronic wavefunctions coupled with the magnetic spin configuration. Such materials can support chiral electronic channels of perfect conduction, and can be used for an array of applications from information storage and control to dissipationless spin and charge transport. Here, we review the theoretical and experimental progress achieved in the field of magnetic topological materials beginning with the theoretical prediction of the Quantum Anomalous Hall Effect without Landau levels, and leading to the recent discoveries of magnetic Weyl semimetals and antiferromagnetic topological insulators. We outline the recent theoretical progress that resulted in the tabulation, for the first time, of all magnetic symmetry group representations and topology. We describe several experiments realizing Chern insulators, Weyl and Dirac magnetic semimetals, and an array of axionic and higher-order topological phases of matter as well as survey future perspectives.

# I. INTRODUCTION

Topological insulators (TIs) and topological semimetals (TSMs), first predicted 15 years ago [1, 2], can exhibit robust boundary states, quantized bulk responses, and exotic transport properties. They represent a possible route towards manipulating quantum information [3], coherent spin transport [4], and high-efficiency catalysis [5]. Whereas a myriad of insulating and (semi)metallic non-magnetic topological phases have now been predicted, characterized and measured, magnetic materials have so far been scarce. These materials' interacting nature renders their theoretical prediction harder than that of their non-magnetic counterparts, yet they are experimentally attractive because magnetism potentially offers greater opportunity for manipulation of topological states. In the past 3 years, theoretical and experimental advances in topological magnetic materials have precipitated [6–11].

For background, in Box 1 we give an overview of the general steps needed for the high-throughput theoretical screening of magnetic topological materials (further technical details are available in Supplementary Information A). Once a candidate topological material has been synthesized (itself a challenge), a variety of experimental tools need to be marshalled to measure their electronic band structures as well as their transport properties and identify topological features: these are summarized in Box 2.

The main purpose of this Review is to discuss recent theoretical and experimental progress in this area using examples from each of the two main material classes – magnetic topological insulators and magnetic topological semimetals. Specifically, we will discuss the characteristic band structure features and transport phenomena of such systems based on two of the most well studied magnetic topological materials: the van der Waals antiferromagnetic $MnBi_2Te_4$ (a magnetic topological insulator) and the Kagome ferromagnetic Weyl $Co_3Sn_2S_2$ (a magnetic topological semimetal). Other magnetic topological possibilities exist that are linked to orbital magnetism instead of spin magnetism; we will defer such discussion to the Supplementary Information B.

Finally, we will conclude with a brief discussion of the opportunities for further theoretical exploration and experimental discovery in this space, and what they might mean for both fundamental studies and practical applications.

## II. THEORY OF MAGNETIC TOPOLOGICAL INSULATORS

The very first magnetic topological insulator is the (integer and fractional) quantum Hall effect, whose discovery and theoretical explanation resulted in two Nobel prizes, in 1985 and 1998. It was subsequently realized [12] by Haldane that an applied magnetic field is not necessary to realize a magnetic topological insulator, and research for the past 10 year theoretically uncovered other insulating phases of matter whose properties are defined by magnetic group topology.

The quantum anomalous Hall effect (QAHE) [12] has been realized by opening a magnetic mass gap (by either magnetic impurities [13, 14] or intrinsic magnetic order [10, 15, 16]) in the Dirac cone at the surface of a thin 3D TI film. This state is, however, the axion (higher order) insulator (AXI) with a chiral hinge mode, where the sample has been thinned to quasi-2D with magnetism added on the surface; it is not the original stoichiometric purely 2D Chern insulator CI [12]. The magnetic order mechanism in these samples is under debate (see [13, 17]), between RKKY (Mn-doped $Bi_2Te_3$ [18]) and Van Vleck [19] (Cr-doped $(Bi,Sb)_2Te_3$ [20]).

The 3D antiferromagnetic (AFM) TI [21, 22] is the only other magnetic TI that has been realized, in $MnBi_2Te_4$ [10] in Type-IV magnetic space group (MSG) $R_I\bar{3}c$ (No. 167.108) [23]. Fig[1] summarizes the different magnetic topological phases observed and predicted in $MnBi_2Te_4$ and related compounds. An AFM TI is the result of turning on the AFM in a nonmagnetic time-reversal (TR)-invariant TI. Forcing doubling of the unit cell, the symmetry group becomes a combination ($\{T|00\frac{1}{2}\}$) of TR (T) and half a lattice translation ($\{00\frac{1}{2}\}$). Before turning on the AFM, the strong TI exhibited a Dirac cone on each surface; the AFM gaps the Dirac cone on the surface perpendicular to the half lattice translation. On the side surfaces, the $\{T|00\frac{1}{2}\}$ enforces Kramers degeneracy at $k_z = 0$ (but not at $k_z = \pi$). The Dirac node sits at $k_z = 0$ in the AFM Brillouin Zone (BZ), *irrespective* of its location (at $k_z = 0$ or $\pi$) in the larger non-magnetic BZ (see Fig. 1), as reported in [10].

Thin samples of MnBi$_2$Te$_4$ produce a QAHE expected to be more robust than that of magnetic-impurity doped TI [24–27]. When a Dirac fermion (exhibiting $\pi$ Berry phase around the Fermi surface) is gapped, the resulting insulator has an integrated Berry curvature - Chern number - $C = 1/2$; the lower and upper surfaces can then either subtract (to lead to $C = 0$) or add, to lead to $C = 1$ QAHE. Whether they add or subtract depends on the surface exchange field. Even/odd number of non-magnetic unit cells will experience opposite/equal exchange field on the top and bottom surface, giving rise to $C = 0/1$. Furthermore, the application of a magnetic field can drive the magnetic moments ferromagnetically. Theoretically, FM MnBi$_2$Te$_4$ (with MSG $R\bar{3}m'$ (No. 166.101)) is a bulk Weyl semimetal (WSM) with one pair of Weyl points (WPs) along the $\Gamma - Z$ [26, 28]. In between the two WPs, in each $k_z$ plane, the Chern number is 1 per $k_z$ momentum. For a thin-film material of $N$ layers, this momentum is quantized in units of $2\pi/N$, which, for $N$ small, gaps the Weyl nodes due to quantum confinement. Hence the finite layer thin film of FM MnBi$_2$Te$_4$ can exhibit a QAHE with a Chern number greater than 1. In [29], a quantum confinement 5 meV gap is induced at the $\Gamma$ point and $C = 2$.

The $C = 1$ QAHE state obtained by gapping a thin film of TI is a *static* AXI, a state of matter with the $\theta$-angle (coefficient of the Chern-Simons form of the Berry potential $\vec{A}(\vec{k})$) equal to $\pi$ [30]. The discovery of higher order TIs (HOTIs) [31] resolved the link between the AXI and the QAHE. T symmetry breaking can drive a 3D TI into an AXI phase with gapped surface states and gapless (chiral) hinge modes [30, 32–38] which carry the Chern number of the QAHE. Inversion-symmetric TIs can be AXIs and at the same time magnetic HOTIs with *intrinsic* hinge states (see Fig. 1). [6, 7, 38–41]. Their bulk topology may be inferred from the Fu-Kane parity formula [32, 33, 42]; *all* of their surface states and Wilson loops are gapped. Gapless hinges provide chiral spectral flow [31, 43].

A large number of theoretically predicted but experimentally undiscovered topological states exist (see Fig.1). While the static AXI has been observed [10, 16, 44–46], it was predicted [47] that a *dynamical* mode of the AXI (phason as the dynamical mode) can arise from gapping *two* Weyl nodes spontaneously by charge-density wave (CDW) formation at the wave-vector between the nodes. This proposal remains unrealized, although its time-reversal counterpart (CDW in a non-centrosymmetric TR invariant Weyl) has been predicted and discovered in (TaSe$_4$)$_2$I [48, 49]. In MSGs, AXI phases can be protected by other bulk symmetries such the product of twofold rotation and T [43, 50, 51] (see Fig.1). The magnetic *Möbius* TCIs [52, 53] host unpaired Dirac-cone surface states – like those of 3D TIs – appearing along surface glide lines and have been predicted in MnBi$_{2n}$Te$_{3n+1}$ [54–56] with canted magnetic moments. They are the T-breaking analogue of the hourglass topological crystalline insulators (TCI) [57, 58] [59] [60] in KHgSb. Glide mirror on some surfaces allows for a degeneracy along mirror symmetric lines in the BZ (see Fig.1).

In total, 4 different topological phases (3 of them shown in Fig.1 e) can be (theoretically) realized in the same compound formula, corresponding to different topological classifications: AFMTI (collinear AFM state), high-order Mobius insulator (in canted AFM state) and mirror TCI (in-plane FM), as well as a FM axion (out-of-plane FM) in MnBi$_8$Te$_{13}$, [61]. The difference in their topological classifications is a clear example of how different magnetic symmetry groups give rise to different topology.

Magnetic topological states implied by symmetry eigenvalues have recently been classified [7, 8], even though material predictions are scarce. Of these, spinful helical magnetic HOTI phases are related to rotational anomalies and exhibit trivial axion angles $\theta = 0(\text{mod } 2\pi)$. When terminated (see Fig.1) in nanorod geometries, the helical magnetic HOTIs generically exhibit even numbers of massless twofold surface Dirac cones (see Fig.1f) on surfaces perpendicular to a rotation axis similar to those in [62]. On their side-surfaces, domain walls between surfaces with oppositely-signed masses bind mirror-protected helical hinge states (Fig.1g).

### III.     MATERIALS FOR MAGNETIC TOPOLOGICAL INSULATORS

There are four main routes for turning a TI into a magnetic QAHE system, sketched in Fig.2: (a) extrinsic deposition of magnetic layers onto the surfaces of the TI, (b) doping the bulk TI with magnetic elements, (c) interleaving magnetic layers into the TI unit cell, (d) identifying intrinsic magnetically ordered TI states. The last two may introduce magnetic symmetries that directly affect the topological classification.

Several attempts made to fabricate hybrid heterostructures of magnetic over-layers on TI surfaces such as EuS over $Bi_2Se_3$ [63–69] have uncovered intriguing transport properties [70], but could not demonstrate quantized anomalous Hall conductance. An alternative path [13], inspired by semiconductor spintronics [71] is doping of the canonical TIs $Bi_2Se_3$, $Bi_2Te_3$ or HgTe with magnetic ions (e.g. V, Mn, Cr, Sm) [13, 68, 72, 73], depicted in Fig.2b. In 2013, the QAHE was measured in thin films of Cr-doped $(Bi_{1-x}Sb_x)_2Te_3$ with a quantized Hall resistance $\rho_{yx}$ observed up to temperatures of 30 mK [13]. In 2015 the QAHE was demonstrated in V doped $(Bi_{1-x}Sb_x)_2Te_3$ with a larger coercive field and in higher temperatures up to 100 mK [73]. While the transport experiments are quite promising, though limited to cryogenic temperatures, the spectroscopic investigation of the energy gap of the corresponding surface states [46, 74–77] has so far been inconclusive. Low Curie temperatures, and the risk of inhomogeneous clustering of dopants thus gave way to intrinsic magnetic TIs.

Recently, a new family of intrinsic AFM TIs was discovered [10, 55, 56, 79] with the general composition $MnTe(Bi_2Te_3)_n$ (Fig.2e and f). $MnBi_2Te_4$ is the first member of the family with a Néel temperature $T_N$ of 25 K [10]. $MnBi_2Te_4$ is a natural heterostructure of MnTe and $Bi_2Te_3$ [10]. The compound topology (band inversion) is akin to the $Bi_2Te_3$ quintuple layer, while the magnetism is related to MnTe. The combined symmetry of time reversal and half a unit cell translation protects the Dirac states parallel to the AFM order from gapping (Fig.2e) but gaps the surfaces perpendicular to it. Spectroscopic reports have been thus far inconclusive on the formation of a magnetic gap at the surface Dirac points: some ARPES measurements find a gapped surface spectrum [10, 80], other image gapless Dirac bands with weak response to lifting of the AFM order above the Neel temperature [81]. Local spectroscopic mappings in STM visualize high level of substitutional Mn atoms on Bi sites, posing a similar challenge to that encountered with magnetically doped TIs [82, 83]. Thin film quantization was shown to give rise to either a QAHE effect, [27] or an AXI [78] (Fig.2g and h, respectively). Accordingly, the $MnTe(Bi_2Te_3)_n$ family will undoubtedly open many new opportunities for magnetic WSMs, and beyond.

Meanwhile, an increasing number of intrinsic magnetic compounds (Fig.2d), are being identified and investigated in transport and spectroscopy in search of clear signatures of broken-TRS induced topology. These include the FM AHE $Fe_3GeTe_2$ [84], the AFM-TI $EuCd_2As_2$ and the AFM-TCI (and possibly a magnetic HOTI) $EuIn_2As_2$, that we briefly discuss. Large anomalous Hall and Nernst signals were detected in the FM $Fe_3GeTe_2$ [84, 85], with $T_c$ higher than room temperature in the few-layer limit of gated devices [86]. The anomalous Hall behavior is believed to originate from the intrinsic Berry curvature contribution due to gapping of a nodal line semimetallic state [84, 87]. No clear spectroscopic evidence of topological states has been provided yet. EuCd$_2$As$_2$ is predicted [88, 89] to turn from a paramagnetic narrow gap semiconductor to AFM-TI below $T_N$=10 K with the easy axis perpendicular to the layers. The combined nonsymmorphic-TRS protects the Dirac surface states on the side surfaces that respect it, while the top and bottom facets are gapped forming a Chern or axion insulating states [88] that have been reported to exist by ARPES [90]. EuIn$_2$As$_2$ is predicted [89] to turn from a paramagnetic narrow gap semiconductor into a type-A AFM Axion insulator [91] or rather a magnetic HOTI [92] below T$_N$=10 K, neither of which is established spectroscopically [93]. A third classification arises when its AFM order aligns parallel to the layers and a magnetic mirror symmetry is restored, classifying the electronic phase as an AFM-TCI.

## IV. THEORY OF MAGNETIC TOPOLOGICAL SEMIMETALS

More than 100 years ago Edwin Hall realized that all FM semimetals and metals exhibit an anomalously large Hall effect (AHE). Since the Hall resistivity versus an applied external magnetic field behaves similarly to the magnetization versus the external magnetic field, it was concluded that the AHE is proportional to the magnetization. Nowadays it is established that the Berry curvature plays an important role in determining the AHE in FM semimetals and metals [94]. Berry noted that an energy-level crossing leads to a physical band crossing that behaves as a magnetic monopole [95], the Weyl point. Magnetic WSMs are common: every crossing point in the band structure of a FM centro-symmetric compound is related to nodal lines or Weyl points.

The simplest topological semimetal, without time-reversal or crystalline symmetry is the solid-state realization of *conventional* Weyl fermions – twofold degeneracies appearing when two singly-degenerate bands cross, at any point in the BZ, and exhibiting linear dispersion away from the degeneracy point [96–98]. Weyl fermions carry a nontrivial topological invariant, the Chern number |$C$| = 1 evaluated on a sphere at energy $E_F$ around the Weyl point.

This invariant renders Weyl nodes locally stable to gapping. The Berry curvature is concentrated near Weyl nodes giving a large AHE (angle) $\sigma_{\vec{W}} = \frac{e^2}{h} C \vec{1}_W$ in magnetic WSMs with Fermi level close to the Weyl nodes [98, 99], where $\vec{1}_W$ is the momentum vector between the two Weyl nodes. By the Mott relation $\alpha_{\vec{W}} = \left(\frac{\pi^3}{3}\right)\left(\frac{k_B^2 T}{e}\right) d\sigma_{\vec{W}}(\epsilon)/d\epsilon_{\epsilon=E_F}$, the ANE is also expected to be large [100–103].

The $\mathbb{Z}$-valued Chern number of the Weyl points reflects the difference in the Chern number of 2D BZ planes above and below the Weyl point. Each BZ plane carrying nonzero Chern numbers projects on surfaces of the crystal to give rise to QH-like edge states, summing up into surface Fermi arcs spanning the momentum space between the projections of the bulk Weyl points. Higher-charge Weyl points appear when two or more Weyl nodes are pinned together by a crystalline rotation symmetry. The first prediction, still unrealized, of a *C* = 2 Weyl node, stabilized by $C_4$ symmetry, was in the FM phase of HgCr$_2$Se$_4$ [104] in the MSG *I*41/*am'd'* (No. 141.557) [23], in which strong AHE was reported [105].

A series of experimentally promising AFM TSM have been predicted [106–108] based on a search of large AHE in Mn$_3$*X* (*X*=Sn, Ge and Ir) and on direct ab-initio calculations [109] in Mn$_3$Sn and Mn$_3$Ge with Kagome layers Mn atoms (see section V 2). The non-collinear magnet Mn$_3$Sn in MSG *Cmc'm'* (No. 63.463) is a magnetic WSM candidate with 6 pairs of Weyl points. Under rigorous MTQC principles [6, 7], it was found that these Weyl points are "accidental": if the 6 Weyl points reported in [109] in half of the BZ were pairwise annihilated without closing a gap at the inversion-invariant momenta, the gapped phase would either be an axion insulator or a 3D QAH state.

CuMnAs and CuMnP have been proposed [110] to exhibit Dirac points. Their AFM order maintains the Type-III symmetry *I*T [23] leading to doubly degenerate bands at each **k** ∈ BZ. Two pairs of these bands cross and their Dirac degenerate point is protected by a non-symmorphic $\{C_{2z}|\frac{1}{2},0,\frac{1}{2}\}$. EuCd$_2$As$_2$ [89] was also proposed as a DSM in a Type-IV MSGs

($D_{3d}^4 \bigoplus T'D_{3d}^4$) [23]. Doubly degenerate bands exist due to $I\{T|0,0,\frac{1}{2}\}$ symmetry, and two pairs can cross with the Dirac point stabilized by $C_{3v}$ symmetry. When three-fold rotation symmetry $C_{3z}$ is broken, the DSM phase can evolve into the AFM TI phase. For magnetic nodal line semimetals were predicted in the layered system $Fe_3GeTe_2$ [87] without SOC. Similarly, in the FM $Co_2MnGa$ [111] with space p group $Fm\bar{3}m$ (No. 225) [23], two majority spin bands near the Fermi level cross on the mirror planes stabilized by mirror symmetry. The nodal lines gap when the SOC is present, although in reality the SOC is negligible. Proposals of Nodal Line Semimetals NLSMs in FM phases of LaCl (LaBr) [112] have not been realized; we believe that most likely these materials are non-magnetic.

Nonmagnetic and magnetic symmetry groups allow 2, 3, 4, 6, and 8-fold degeneracy "new fermions" [113, 114] in the BZ. In (Type I [23]) space groups 3-, 4- and 6- dimensional degeneracies can appear; Type-III and Type-IV groups [23] support [113, 114] 8-fold double Dirac points [115] degeneracies. The chiral AFM phase of $Mn_3IrSi$ is predicted [114] to host Spin-1 Weyl fermions with 3-fold degeneracies. $Mn_3IrSi$ [114] and $Nd_5Si_4$ [6] are predicted to be chiral magnetic TSM. The 4, 6, and 8-fold new degeneracies are not protected by a Chern number (as in the case of WSM) and hence do not exhibit Fermi arcs on surfaces; they are novel Higher Order Semimetals (HOTSM) exhibiting *hinge* arcs [116]. A simple model for hinge arcs can be expressed as a $k_z$ phase transition between a Quadrupole Insulator (QI) in [31] and a trivial insulator. Related arguments show that both Dirac HOTSMs and 6-fold degeneracies *universally* host intrinsic hinge states [116, 117].

### v.  MATERIALS FOR MAGNETIC TOPOLOGICAL SEMIMETALS

*1.Ferromagnetic compounds*

The FM WSM $Co_3Sn_2S_2$ in MSG $R\bar{3}m'$(No. 166.101) has been extensively explored and characterized for its topological properties. Its crystal structure is composed of A-B stacked triangular layers of Sn and S and Kagome layers of magnetic Co ions (Fig.3, inset) captured in STM topography [9]. The compound hosts one electron more than the semiconducting non-magnetic Shandite $Co_3InSnS_2$. $Co_3Sn_2S_2$ fully polarized spin ($0.29\mu_B$/Co) leads to a half metallic FM with a relative high Curie temperature of 177 K with its spins oriented out of plane [119]. A single valence and conduction band cross the Fermi energy, leading to prediction of Weyl crossings, shown in Fig.3b. Experimental evidence for the bulk Weyl nodes close to the Fermi energy was provided by ARPES measurements (Fig.3c) [118], as well as confirmed by STM through quasi particle interference (Fig.3d).

Clear magneto-transport signatures of the magnetic topological state were reported prior to the spectroscopic verification. These include negative magneto-resistance under parallel current and magnetic field (Fig.3g), potentially signifying chiral anomaly, high anomalous Hall conductivity [119], and a significantly higher anomalous Nernst signal than conventional materials (Fig.3f) [120]. STM further finds presence of linearly dispersing step-edge modes (Fig.3g) [121], while theory predicts isolated $Co_3Sn$ sheets will exhibit QAHE [123]. Furthermore, the Kagome structure of the magnetic Co ions can host flat band models due to the line-graph property of the lattice [124]. Intriguingly, a zero bias conductance peak has been detected in STM (Fig.3h) on the Co surface termination, with an unusual response to magnetic field [122, 125]. Bulk single crystals of $Co_3Sn_2S_2$ have been even used for a proof-of-concept investigations of the efficiency towards water oxidation [126].

All these suggest a new direction to search and synthesize magnetic TSMs in Kagome [109, 119, 127] and honeycomb-layer of a 3$d$-transition metal ions [128]. This family of materials exhibit Weyl and Dirac fermions in both FM and AFM materials. Examples to date include: FeSn [129], $Fe_3Sn_2$ [130], $Mn_3Sn$ [131], $Mn_3Ge$ [106, 132], and CoSn [133], as well as the $R$Mn$_6$Sn$_6$ family with $R$ = Tb, Gd, Tm, Lu [134–136]. Experimental signatures include a temperature independent enhanced AHE up to room temperature in $Fe_3Sn_2$ [130], and a gapped two-dimensional Dirac band close to the Fermi energy by ARPES. A giant spontaneous nematic energy shift, larger than any possible Zeeman splitting, hints at strong correlations in $Fe_2Sn_3$ [137]. To reduce the dimensionality and increase correlations effects, FeSn - with decoupled iron layers - was identified as an ideal Kagome lattice [129]. Flat bands and fully spin-polarized surface states (confirmed by ARPES) suggest the presence of spatially decoupled Kagome planes. For the summery of the anomalous transport properties of the Kagome compounds and the corresponding synthesis methods, please see Table I.

Another large family of half metallic $Co_2$-Heusler compounds holds great potential as magnetic Weyl candidate materials because of their tunability. They were proposed in $Co_2YZ$ ($Y$ = V, Zr, Nb, Ti, Hf, $Z$ = Si, Ge, Sn) [138] and $Co_2MnZ$ ($Z$ = Ga, Al) [139] and can be grown in bulk and thin films [140]. A leading magnetic topological semimetal candidate is $Co_2MnGa$ that was verified spectroscopically in ARPES as a nodal line semimetal hosting drum head surface states [111]. Strong AHE indicating interplay between nodal lines and their partial gapping into Weyl points [141] close to the Fermi energy were reported in it and in $Co_2MnAl$ [139]. The AHE in $Co_2MnGa$ and $Co_2MnAl$ is even larger than that magnetically induced in GdPtBi, leading to a Hall angle of 12 % [142] and 21 % [141], respectively. The synthesis methods, and the transport properties of GdPtBi and $Co_2MnZ$ are summarized in Table I. Additionally, the large ANE signal is achievable at lower magnetic fields as it scales beyond the magnetization due to the Berry phase contribution. In $Co_2MnGa$ an ANE with a remarkably high value $SA_{yx}$ of ∼6.0 $\mu_V K^{-1}$ at room temperature, (an order of magnitude higher than for conventional FMs), was reported [102, 143, 144]. High-throughput searches for large Berry phase contributions close to the Fermi energy have identified several magnetic compounds with the naturally abundant and low-cost element iron, such as the nodal line compounds $Fe_3Ga$ and $Fe_3Al$ [145]. The expected high efficiency of lateral thin film devices may pave the way for new large-area energy harvesting technology.

## 2. Antiferromagnetic compounds

The first magnetically induced WSM was realized in the AFM half Heusler compound GdPtBi [146]. GdPtBi and NdPtBi become WSMs only in applied fields of the order of 2 T [146–148]. Strong signatures typical for magnetic WSMs were observed in both compounds including the chiral anomaly, the gravitational anomaly, a large non-saturated negative quadratic magnetoresistance for fields of up to 60 T, an unusual intrinsic anomalous Hall effect, and planar Hall effect [146, 148–150]. In most AFM compounds the magnetic ordering is unknown, because large single crystals are needed for neutron scattering. Considering various AFM orders allows to construct a generalized Kane model with a resulting rich phase diagram ranging from Dirac, Weyl and nodal semimetal phases, type-B triple point phases, topological mirror (or glide), and AFM topological insulating phase [151]. The topological nature of Heusler half metallic compounds such as MnPtSb and MnPtBi, with a Curie temperature up to 1000 K [140], is still unexplored.

AFM order offers an even richer magnetic phase diagram than FM order. Collinear antiferromagnets with a zero net magnetic moment must have a net zero Berry curvature, although crossings can be sometimes observed in the band structure of an AFM metal. In agreement with this simplified picture the AHE is absent in nearly all AFMs that have zero magnetization. The three systems, hexagonal $Mn_3Sn$ [131], $Mn_3Ge$ [106, 132] and cubic $Mn_3Ir$ [107] have non-collinear triangular AFM arrangements, which is the origin of a non-vanishing Berry curvature. $Mn_3Sn$ and $Mn_3Ge$ have Weyl points close to the Fermi energy, and show the predicted properties of an AHE even at room temperature [131, 132] and exhibit complex Fermi arcs in qualitative agreement with theory [131]. The chiral anomaly, was also reported in the anisotropic compound $Mn_3Sn$ [152] as was a large ANE [153] and magneto optical Kerr effect [154]. The strong response of the WSM compounds to external stimuli makes them promising candidates for topological AFM spintronics [155]. Another family of compounds exhibiting an intricate AFM order and showing promising experimental signatures in spectroscopy and magnetotransport, such as a singular angular magnetoresistance (SAMR), exceeding 1000% per radian, is $R$AlGe with $R$ = Ce [156, 157], Pr [158] and the nonmagnetic La [159]. Lastly, we remark that the pyrochlore irridate family of materials, in which topological WSMs were first predicted [97], has remained largely unexplored. Their tetrahedral spin configuration gives rise to rich AFM orders [160] coupled to magnetic topological phases [161–163] as well as correlated AFM Mott states [164–166]. Remarkably, metallic modes [167] on AFM domain walls, thought to be a precursor of Fermi-arc states [168], were imaged within the otherwise Mott insulating state.

## VI. FUTURE DIRECTIONS

Using high-throughput searches [6] a more systematic search for magnetic topological materials with high Curie temperatures, is important for quantum (computing/sensors) and classical (thermoelectrics/Hall sensors/efficient catalysts) applications. These following additional characteristics should be considered for material selection: (a) topological magnets with high Curie temperatures, which enable a high ANE signal close to room temperature; (b) low magnetic moment for eliminating stray magnetic fields in devices; or (c) hard magnets favorable for AHE and ANE at zero magnetic field; (d) low dimensional crystal structure and electronic structures for quantum confinement; and, finally, (e) frustrated atomic arrangement such as Kagome lattices for flat bands and non-collinear spin structures.

The design of a material that exhibits a high temperature QAHE via quantum confinement of a magnetic WSM, and its integration into quantum devices is desired. Indeed, several magnetic topological semimetals and insulators are predicted to realize QAHE in the thin film limit [84, 123, 172]. The realization of the QAHE at room temperature would be revolutionary, overcoming the limitations of data-based technologies, which are affected by large electron scattering-induced power losses. This would pave the way to new generations of low energy consuming quantum electronic and spintronic devices.

Magnetic topological systems are a fertile field for further theoretical discoveries. While the complete stable topological indices of magnetic and nonmagnetic TIs, TCIs, and TSMs have been computed [6–8, 40, 173], the magnetic fragile topological indices remain an outstanding problem [174]. A classification of the magnetic obstructed atomic insulators - phases of matter described by bands which are not topological in the sense that they admit a localized Wannier description, but whose Wannier centers do not locate at the atom positions [175] – is still outstanding.

A further fundamental breakthrough would be the development of a framework to predict crossing points of WSM or DSM that are *very close* to the Fermi energy. The studying of topological materials displaying incommensurate magnetism is almost non-existent at the present time and should be pursued; it is unclear if in this case other topological classes besides the Chern QAHE class exists.

In parallel, the magnetic topological responses must also be developed. While for non-magnetic systems, we understand a series of responses such as chiral and gravitational anomaly, quantized photo-galvanic effects in Weyl, and topological defects we do not yet understand any *specific* responses that are unique to magnetic systems. A full classification of magnetic topological defects in crystalline TIs is absolutely necessary. Predictions of magneto-optical responses are needed, especially in the new, rotation-anomaly magnetic TIs. Predictions of *quantized* responses are particularly desired.

The field of magnetic topological superconductivity is also completely open. Since magnetic materials already have time-reversal breaking, they might exhibit topological excitations such as Majorana zero modes without the need for applying a magnetic field, thereby rendering these systems useful for practical applications. While the types of phases that one obtains by proximitizing the surface of non-magnetic topological systems is mature, the similar studies for all the magnetic crystalline TIs are absent.

The next step is the theoretical introduction of bulk/surface interactions in topological materials. In the bulk, topological metals can give rise to many-body states by tuning interactions. The simplest example is a Weyl CDW axionic insulator, mirroring the non-magnetic experiment [48]. Many-body effects on the surface of magnetic TIs could give rise to topologically ordered states of matter.

**ACKNOWLEDGMENT**

B.A.B. work on magnetic topology is mainly supported by the DOE Grant No. DE-SC0016239. Further support comes from the Schmidt Fund for Innovative Research, Simons Investigator Grant No. 404513, the Packard Foundation, the Gordon and Betty Moore Foundation through Grant No. GBMF8685 towards the Princeton theory program, the NSF-EAGER No. DMR 1643312, NSF-MRSEC No. DMR-1420541 and DMR2011750, ONR No. N00014-20-1-2303, Gordon and Betty Moore Foundation through Grant GBMF8685 towards the Princeton theory program, BSF Israel US foundation No. 2018226, and the Princeton Global Network Funds. C. F. was supported by the ERC Advanced Grant No. 742068 'TOPMAT' and by the Deutsche Forschungsgemeinschaft (DFG, German Research Foundation) under Germany's Excellence Strategy through Würzburg-Dresden Cluster of Excellence on Complexity and Topology in Quantum Matter - ct.qmat (EXC 2147, project-id 390858490). B. A. B. received additional support from the Max Planck Society. Additional support was provided by the Gordon and Betty Moore Foundation through Grant GBMF8685 towards the Princeton theory program. H.B. acknowledges support from the European Research Council (ERC) under the European Union's Horizon 2020 research and innovation program (grant agreement no. 678702), and the German-Israeli Foundation (GIF, I-1364-303.7/2016).

Zahid Hasan, Discovery of topological Weyl fermion lines and drumhead surface states in a room temperature magnet, Science **365**, 1278 (2019).

**This is the first proof of a ferromagnetic nodal line half metal with surface states that take the form of drumheads via ARPES in Co$_2$MnGa.**

# Figures

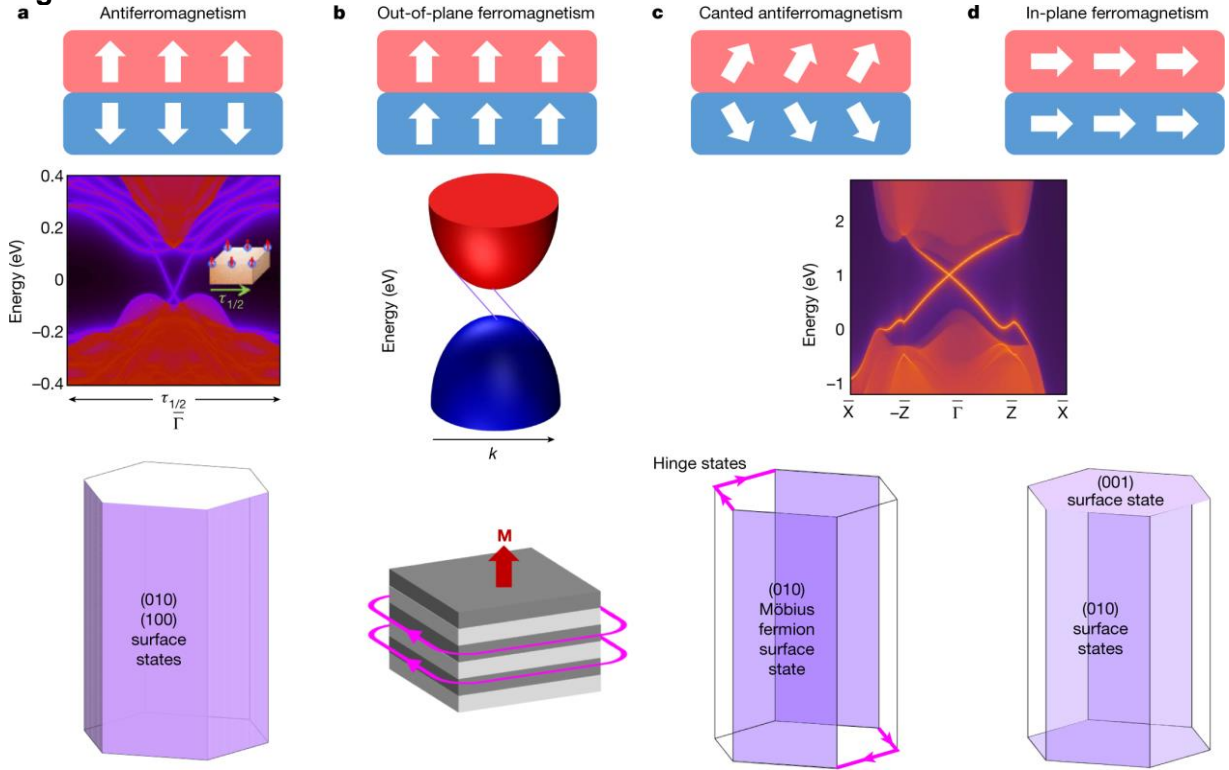

FIG. 1. **Interplay between magnetic orders and topology.** Depending on the spin configuration, the MnBi$_{2n}$Te$_{3n+1}$ system is predicted to be (a) an AFMTI - with a single gapless Dirac cone protected by $\{T|00\frac{1}{2}\}$ on the symmetry-preserving (010) (or (100) surface), while the symmetry non-preserving (001) surface is gapped, (b) in a thin 2D sample with only a few layers, a QAHE state with $C = 1$ (AXI) or $C = 2$ depending on the number of layers and with Chiral edge states (c) a Möbius Insulator in canted AFM which respects glide mirror $\{M_x|00\frac{1}{2}\}$ symmetry with $M_x$ mirror followed by half lattice translation. The insulator shows surface states on the symmetry preserving (010) surface but not on the (100) and (001). Two opposite (010) surfaces are linked together by 1D chiral hinge states, manifesting the higher-order nature (HOTI) of the system. The surface state is a Dirac cone, whose position is on the $\bar{\Gamma} - \bar{Z}$ line, and their mirror eigenvalues, proportional to $e^{ikz/2}$ require two BZ ($4\pi$) to return to themselves, hence the Möbius name. (d) a TCI phase for in-plane FM, where the glide mirror $\{M_x|00\frac{1}{2}\}$ is promoted to a mirror $M_x$, and now a surface state appears on both symmetry preserving (010) and hte (001) surfaces.

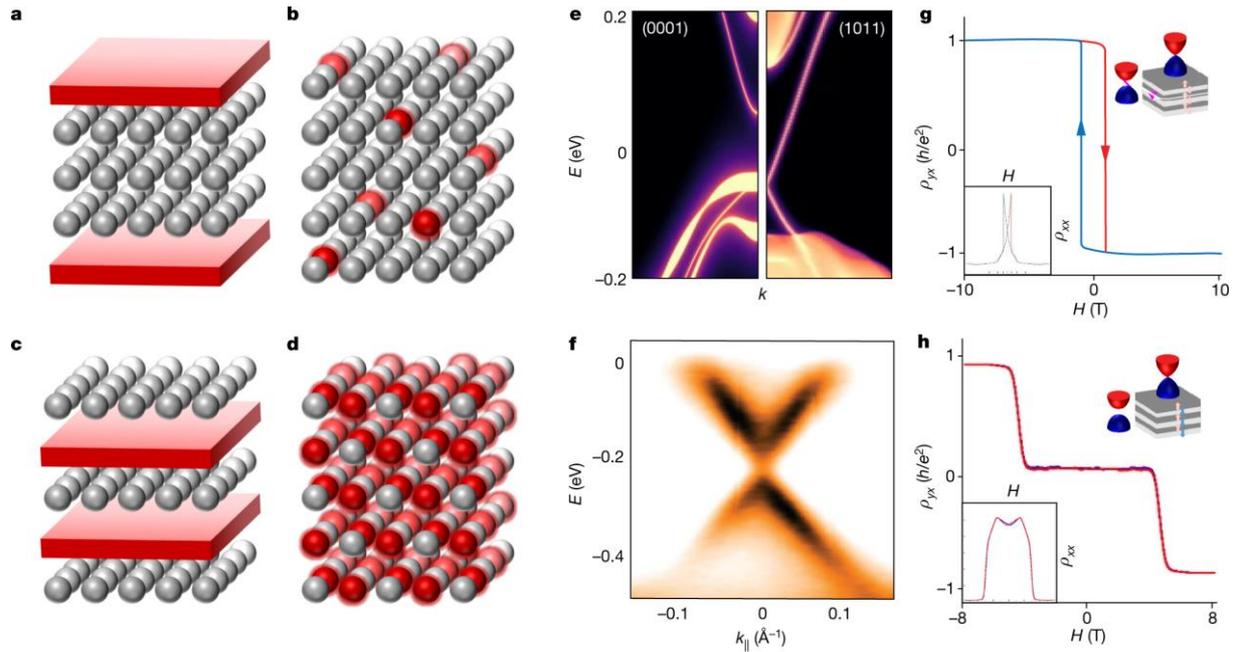

FIG. 2. **Magnetic topological insulators - Realizations and MnBi$_2$Te$_4$ case study.** There are four main approaches for realizing a magnetic TI: (a) Deposition of a magnetic layer over a TI surface (eg EuS/Bi$_2$Se$_3$). (b) Magnetic element doping of a TI (e.g. Bi$_{2-x}$Cr$_x$Te$_3$). (c) Stoichiometrically interleaved magnetic layers within the TI unit cell (e.g. MnBi$_2$Te$_4$). (d) Intrinsically stoichiometric magnetic TIs (e.g. EuCd$_2$As$_2$). The AFM TI MnBi$_2$Te$_4$ was recently shown to exhibit quantized anomalous Hall conductance in the thin film limit with an exciting dependence on parity of layer number (e) ab initio calculation of its surface band structure predicts presence of Dirac states on surfaces that preserve combined time reversal and half unit cell translation operation and massive ones on those that that break it (right and left panels, respectively) [28]. (f) spectroscopic ARPES measurement images Dirac surface states with possible induced gap at the Dirac node [10]. The magneto-transport responses show prominent magnetic topological response (g) For even septuple-layer thin film QAHE is observed with quantized Hall resistance [15] (h) while for odd septuple-layer axion response is found with null Hall response [78].

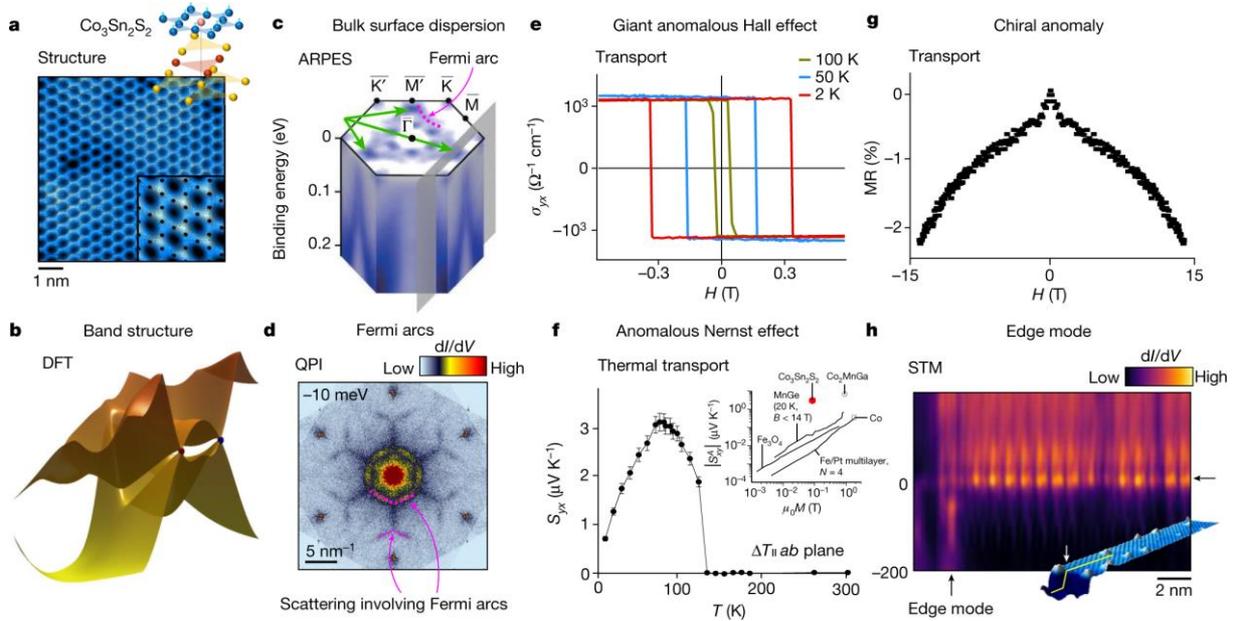

FIG. 3. Magnetic Topological semimetals - $Co_2Sn_2S_2$ case study The ferromagnetic WSM $Co_2Sn_2S_2$ exhibits many of the phenomena associated with magnetic topological matter (see Box 2 (a) Atomic structure contains Kagome layers of Co magnetic ions (inset) seen in STM topography [9]. (b) ab initio calculation of its band structure within the FM phase finds 6 bulk Weyl nodes with Dirac-like dispersion. (c) The bulk Weyl bands and corresponding surface Fermi arc states (magenta) are captured in ARPES [118] (d) Scattering processes involving Fermi arc states are also imaged in STM through QPI (magenta) [9]. (e) Transverse magnetotransport ($B\perp I$) finds a giant AHE [119] (f) as well as large anomalous Nernst signal ($B\perp \nabla T$) that onsets with finite coercive field within the FM phase [120]. (g) Strong negative magneto resistance ($B\|I$) is further detected possibly signifying chiral anomaly [119]. (h) STM spectroscopic imaging finds an increased density of states (marked by open arrow) localized next to crystallographic step edge (topography in inset) attributed to edge mode [121]. Zero-bias conductance peak (marked by solid arrow) possibly originates from geometric frustration due to Kagome structure [122]

**BOX 1. Prediction of magnetic topological materials. WaveFunction Representations.**
The framework for classifying and analyzing magnetic crystals and their topology is that of the 1651 magnetic symmetry groups (MSGs). A complete catalogue of MSGs and their topological phases is now available, for the first time, on the Bilbao Crystallographic Server (BCS). The steps for high-throughput material discovery are: First, an ab-initio calculation obtains the electronic wavefunction representations (called "symmetry data") at high symmetry points for a specific material. Efficient M/TQC [176–178] and related methods [40, 173, 179, 180] require only the electronic wavefunctions at a small number of specific, high-symmetry momenta. **Topological Indices.** Second, the symmetry data is used to compute the topological indices of a set of bands. The computation background is rather complicated and tedious for a large number of bands; publicly available codes (Check Magnetic Topology) can now automate this process. Given the information available in the band's symmetry data, the code checks both whether a set of bands in the Brillouin zone can form an insulator. If not, they are a TSM. If yes, the code further checks whether the insulator can be expressed as a sum of atomic, trivial insulators. If not, it is topological. Further division of topological class is performed and materials are tabulated in a way similar to the spirit of a "periodic table" of compounds. **Material Prediction.** Band structures and topological phase diagrams depending on Hubbard $U$ interaction are computed and posted in Topological Materials Database. Despite the results of [6–8], and the recent experimental identification of novel magnetic TIs [10, 16] and TSMs [9, 111, 118], the low number of materials with magnetic structures measured through neutron diffraction and matched with magnetic SGs hinders magnetic material discovery. Neutron scattering measurements [181] for every magnetic compound, coupled with theoretical analysis of the spin wave spectrum are needed to determine the MSG. Fortunately, the list of compounds for which the MSG is determined by the MAGNDATA application on the BCS [182] is growing daily, from 500 a year ago [6] to 1000 today. Knowing the MSG allows for far more accurate ab-initio calculations of the band structure, which can then be used [7] to determine the topological classification of the magnetic materials.

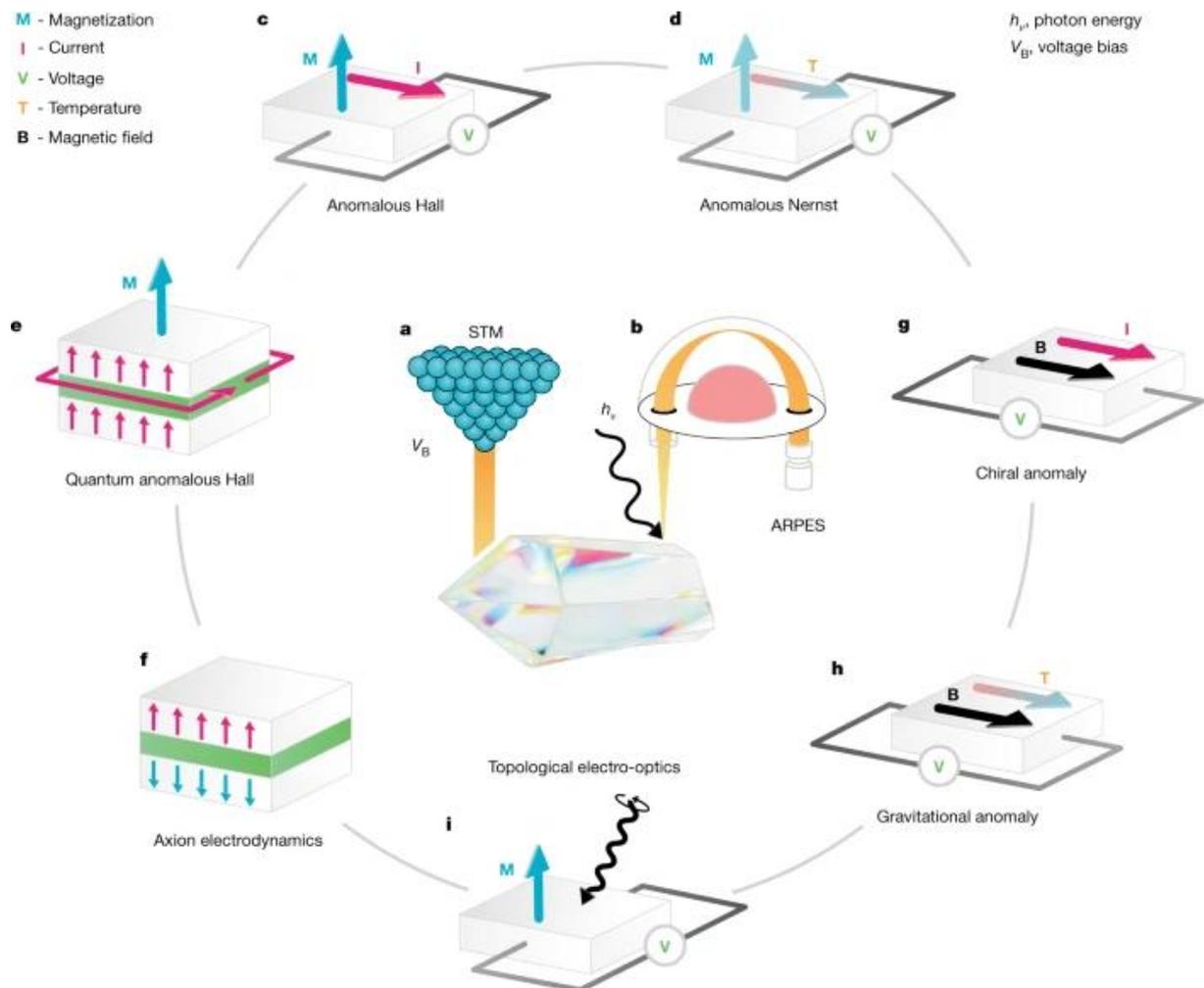

**BOX 2. Experimental identification of magnetic topological materials** Band structure mapping: a) Scanning tunneling microscopy (STM) and b) Angle-resolved Photoemission Spectroscopy (ARPES) probe spectroscopic topological fingerprints mainly through the concept of bulk-boundary correspondence of the electronic band structure. Magnetic Weyl and Dirac semimetals exhibit linear dispersion in the bulk and Fermi arcs at their surfaces, whereas nodal line semimetals host complex drum head surface states [9, 111, 118, 183]. In magnetic TIs the main spectroscopic challenge is to image the gapped surface states [27, 28]. Transport and optical properties: Magnetic topological materials that host an enhanced Berry curvature further exhibit extreme responses to external stimuli such as magnetic field, voltage or current bias, temperature gradient and optical excitation that can be applied in various longitudinal and transverse combinations as shown in c) to i) [102, 119, 120, 143, 153]. c) For observing the anomalous Hall effect, an electric current is injected normal to the magnetization. The resulting Hall resistivity of a ferro- or ferrimagnetic compound is then proportional to the magnetization and enhanced close to Dirac and Weyl points or nodal lines [94] d) Analogous to the AHE, an ANE is observed in measurements of heat current, where a transverse voltage is produced by a temperature gradient and the magnetization orthogonal to each other. e) In extreme cases the anomalous Hall response becomes quantized [27] f). Whenever the magnetization on opposite boundaries has opposite polarity an Axion insulator is formed in which the electrons exhibit axion electrodynamics. g) The chiral anomaly in topological materials is manifested by a negative magneto-resistance in response to parallel electric current and magnetic field. The negative magneto resistance arises from the magnetic-field-induced imbalance in the number of fermions in each pair of Weyl nodes with opposite chirality. h) The same experimental setup with a thermal gradient instead of an electric field, probes the gravitational anomaly. I) Nonlinear optical conductivity components should be enhanced due to the large Berry in semimetal and nodal line compounds.

TABLE I. Synthesis-methods, Anomalous transport properties of magnetic topological semimetals.

| Compound | Synthesis | Topology type | Magnetism | $T_C$ or $T_N$ (K) | AHC (~2K) ($\Omega^{-1} cm^{-1}$) | ANE ($\mu V K^{-1}$) | Ref. |
|---|---|---|---|---|---|---|---|
| GdPtBi | Flux | Weyl | AFM | 9 | 200 | – | [169] |
| $Co_2MnGa$ | Bridgeman | Nodal line | FM | 686 | 1600 | 6 (300 K) | [102, 143, 144] |
| $Co_2MnAl$ | Floating-zone | Nodal line | FM | 726 | 2000 | – | [141, 170] |
| $Co_3Sn_2S_2$ | Flux, CVT | Weyl | FM, non-collinear | 177 | 1130 | 3 (80 K) | [119, 120] |
| FeSn | Flux | Dirac | AFM | 365 | – | – | [129] |
| $Fe_3Sn_2$ | CVT | Dirac | FM | 670 | 1050 | – | [130] |
| $Fe_3Ga$ | Czochralski | Nodal line | FM | 720 | 610 | 6 (300 K) | [145] |
| $Fe_3Al$ | Czochralski | Nodal line | FM | 600 | 460 | 4 (300 K) | [145] |
| $Fe_3GeTe_2$ | CVT | Nodal line | FM | 220 | 540 (10 K) | 0.3 (50 K) | [122] |
| $Mn_3Sn$ | Bridgeman, Czochralski | Weyl | AFM, non-collinear | 100 | 100 (100 K) | 0.6 (200 K) | [131, 153] |
| $Mn_3Ge$ | Czochralski | Weyl | AFM, non-collinear | 500 | 1.5 (100 K) | – | [132, 171] |